\documentstyle[twoside,fleqn,espcrc2]{article}

\begin{document}

\input{psfig}

\bibliographystyle{unsrt}

\newcommand{\ltsim}{\displaystyle\mathop{<}_{\sim}}

\hyphenation{bet-ween con-trac-tions simu-la-tion con-figu-ra-tion se-ve-ral
Ho-shi-no Shi-ra-ka-wa Oya-na-gi Ichii Ka-wai Tsu-ku-ba}

 
\title{Light hadron masses with an $O(a^2)$ improved NNN action
\thanks{Supported in part by the Natural Sciences and Engineering
Research Council of Canada, and by NSF PHY-9409195.}}

\author{H.R.~Fiebig\address{Physics Department, F.I.U.-University Park,
Miami, Florida 33199, U.S.A.} and R.M.~Woloshyn \address{TRIUMF, 4004
Wesbrook Mall, Vancouver, B.C. Canada, V6T 2A3}}
       
\begin{abstract}
Meson and baryon masses in the light (u,d and s) sector are calculated
using tadpole-improved gauge field and fermion actions. These are corrected
to order $O(a^2)$ on the classical level using next-nearest-neighbour terms.
The results, obtained at lattice spacings of 0.4 and 0.27fm, are compared
to Wilson action calculations.
\end{abstract}
\maketitle

Improved actions hold great promise for moving lattice field theory closer to
phenomenology. Additional terms in the action that are nonleading in powers of
the lattice spacing $a$ can be used to reduce discretization errors. These
can be significantly suppressed further by incorporating tadpole factors
\cite{lep_mac} with the nonleading terms in the action.

It is desirable to push improvement to the next-to-leading order in $a$.
Alford et al proposed the so-called D234 action \cite{alfor_coars}.
This action is corrected to $O(a^2)$ and tadpole improved. It is
built by adding clover \cite{sw} and next-nearest-neighbour terms to the
Wilson action.

A simpler next-nearest-neighbour (NNN) action was considered
some time ago \cite{nnn}. Tree-level $O(a^2)$ improvement is achieved by
employing 2-link terms for the fermionic part and 6-link rectangles
for the gauge field. We here use this NNN
action with tadpole improvement to compute light hadron masses.
Calculations with this 
action were also reported by Bock \cite{Bock} and by Bori{\c{c}}i and 
de Forcrand \cite{Borici}.

The $SU(3)$ gauge field part of the action is
\begin{eqnarray}
S_G(U) &=& \beta\left[\sum_{pl}(1-\frac13\mbox{Re}\mbox{Tr}U_{pl})
\right.\nonumber \\ & & \hspace{7mm}
\left. +C_{rt}\sum_{rt}(1-\frac13\mbox{Re}\mbox{Tr}U_{rt})\right] \,.
\label{eq1}\end{eqnarray}
The first term is the 4-link plaquette (pl) Wilson action and $U_{rt}$ are
the planar 6-link rectangles (rt) which are sufficient to remove $O(a^2)$
errors at the classical level. The coefficient $C_{rt} = -1/20U_0^2$
includes the tadpole factor
\begin{equation}
U_0 = \langle\frac13\mbox{Re}\mbox{Tr}U_{pl}\rangle^{1/4} \,.
\label{eq2}\end{equation}
For the fermions  NNN couplings in both the kinetic and Wilson
terms are used \cite{nnn}
\begin{eqnarray}
S_F(\bar{\psi},\psi;U) &=& \nonumber \\
& & \hspace{-25mm} \phantom{-}\sum_{x,\mu}\frac{4}{3}\kappa\left[
\bar{\psi}(x)(1-\gamma_{\mu})U_{\mu}(x)\psi(x+\mu)\right. \nonumber \\
& & \hspace{-4mm} \raisebox{1.2ex}[1ex][0ex]{$ \left. +\bar{\psi}(x+\mu)
(1+\gamma_{\mu}) U_{\mu}^{\dagger}(x)\psi(x)\right] $} \nonumber \\
& & \hspace{-25mm} -\sum_{x,\mu}\frac{1}{6}\frac{\kappa}{U_0}
\left[\bar{\psi}(x)(2-\gamma_{\mu})U_{\mu}(x)\right. \nonumber \\
& & \hspace{11mm} \raisebox{1.4ex}[1ex][0ex]{$
U_{\mu}(x+\mu)\psi(x+2\mu) $} \nonumber \\
& & \hspace{-21mm} \left.+\bar{\psi}(x+2\mu)(2+\gamma_{\mu})
U_{\mu}^{\dagger}(x+\mu)U_{\mu}^{\dagger}(x)\psi(x)\right] \nonumber \\
& & \hspace{-25mm} -\sum_{x}\bar{\psi}(x)\psi(x).
\label{eq3}\end{eqnarray}
The Wilson action is recovered by replacing the coefficient $4/3$ by $1$
and dropping NNN terms.

\begin{table*}[htb]
\newlength{\digitwidth} \settowidth{\digitwidth}{\rm 0}
\catcode`?=\active \def?{\kern\digitwidth}
\caption{Lattice Details. $N_U$ is the number of gauge configurations and
$a_{st}$ is the lattice spacing determined from the string tension
\protect\cite{hdt}. $\kappa_s$ is the hopping parameter corresponding to
the strange quark mass.}
\label{tab1}
\begin{tabular*}{\textwidth}{@{}l@{\extracolsep{\fill}}ccccccc}
\hline
Action & Lattice & $N_U$ & $\beta$ & $a_{st}$[fm] &
 $\kappa$ & $\kappa_s$ \\ \hline
NNN & $6^3\times 12$ & 160 & 6.25 & 0.4 &
 0.162, 0.165, 0.168, 0.171, 0.174 & 0.166  \\
NNN & $8^3\times 14$ &  60 & 6.8  & 0.27 &
 0.148, 0.150, 0.152, 0.154, 0.156, 0.158 & 0.1558 \\
Wilson   & $6^3\times 12$ & 160 & 4.5  & 0.4 &
 0.189, 0.193, 0.197, 0.201, 0.205, 0.209, 0.213 & 0.205 \\
Wilson   & $8^3\times 14$ &  90 & 5.5  & 0.27 &
 0.164, 0.168, 0.172, 0.176, 0.180 & 0.178 \\ \hline
\end{tabular*}
\end{table*}

\begin{table*}[ht]
\caption{Results of the calculations extrapolated to the limit $M_{\pi}=0$.
The quantity $J$ is defined in \protect\nocite{smear,lanl}\protect\cite{lm}.}
\label{tab2}
\begin{tabular*}{\textwidth}{@{}l@{\extracolsep{\fill}}cccccc}
\hline
   & \multicolumn{2}{c}{Improved} & \multicolumn{2}{c}{Wilson} & Exp. \\
   & $\beta=6.25$ & $\beta=6.8$ & $\beta=4.5$ & $\beta=5.5$ \\ \hline
$M_{\rho}a_{\rho}$ & 1.19(5) & 0.90(5)         & 0.90(2) & 0.71(3) \\
$a_{\rho}^{-1}$ & 648(27)MeV & 855(45)MeV & 858(15)MeV & 1085(46)MeV \\
$a_{st}/a_{\rho}$ & 1.31(5) & 1.17(7)    & 1.73(3) & 1.50(6) &      \\
$J$               & 0.43(8) & 0.38(7)    & 0.31(2) & 0.32(7) &      \\
$K/{\rho}$        & 0.65(2) & 0.65(4)    & 0.61(1) & 0.61(4) & 0.64 \\
$K^{\ast}/{\rho}$ & 1.17(3) & 1.16(4)    & 1.10(1) & 1.13(4) & 1.16 \\
$\phi/{\rho}$     & 1.31(4) & 1.30(6)    & 1.20(2) & 1.24(6) & 1.32 \\
$N/{\rho}$        & 1.55(6) & 1.36(9)    & 2.05(5) & 1.73(14)& 1.22 \\
$\Delta/N$        & 1.34(4) & 1.38(11)   & 1.07(2) & 1.24(10)& 1.31 \\
$\Sigma/N$        & 1.15(2) & 1.20(4)    & 1.05(1) & 1.10(5) & 1.27 \\
$\Xi/N$           & 1.23(3) & 1.32(4)    & 1.09(1) & 1.15(7) & 1.40 \\
$\Lambda/N$       &         & 1.17(4)    &         & 1.08(5) & 1.19 \\
$\Omega^-/N$      & 1.58(4) & 1.67(10)   & 1.19(2) & 1.38(10)& 1.78 \\ \hline
\end{tabular*}
\end{table*}

\begin{figure}[htb]
\vspace{9pt}
\makebox[75mm]{\psfig{figure=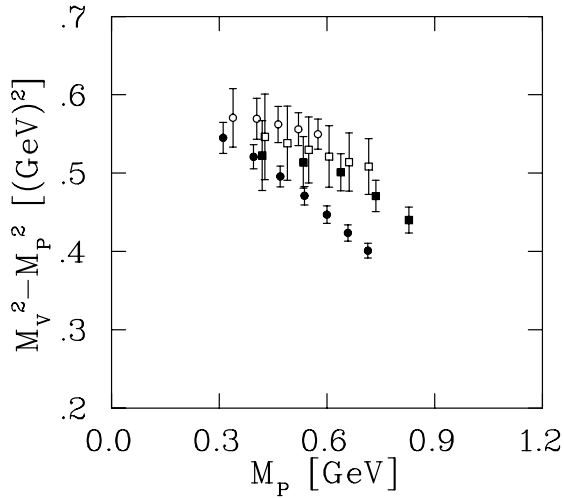,height=65mm}}
\caption{The vector-pseudoscalar squared mass difference for the Wilson
(solid symbols) and the NNN (open symbols) action for the $\beta$ values
of Tab.~\protect\ref{tab1}.}
\label{fig1}
\end{figure}
 
\begin{figure}[htb]
\vspace{9pt}
\makebox[75mm]{\psfig{figure=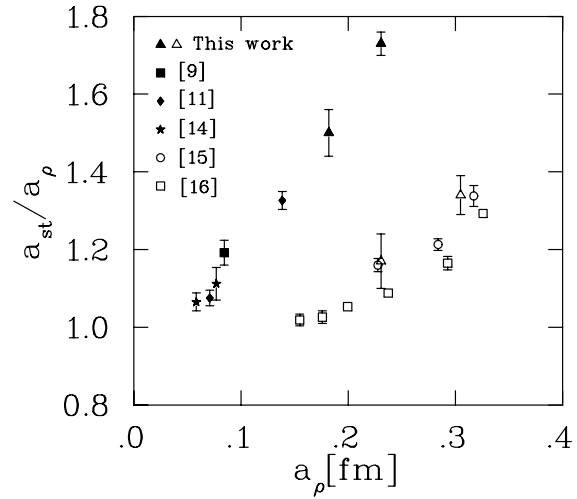,height=65mm}}
\caption{Ratio of the lattice spacings determined by the string tension
and the $\rho$-meson mass versus $a_{\rho}$ for the Wilson (solid symbols)
and improved actions (open symbols).}
\label{fig2}
\end{figure}

\begin{figure}[htb]
\vspace{9pt}
\makebox[75mm]{\psfig{figure=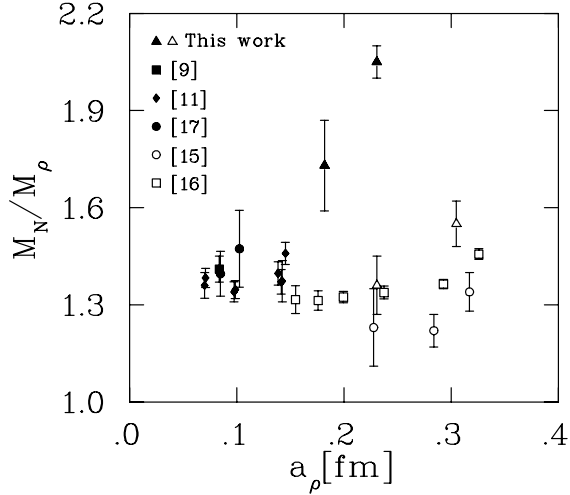,height=65mm}}

\caption{Ratio $M_N/M_{\rho}$ versus $a_{\rho}$ for improved actions (open
symbols) and the Wilson action (solid symbols).}
\label{fig3}
\end{figure}

\begin{figure}[htb]
\vspace{9pt}
\makebox[75mm]{\psfig{figure=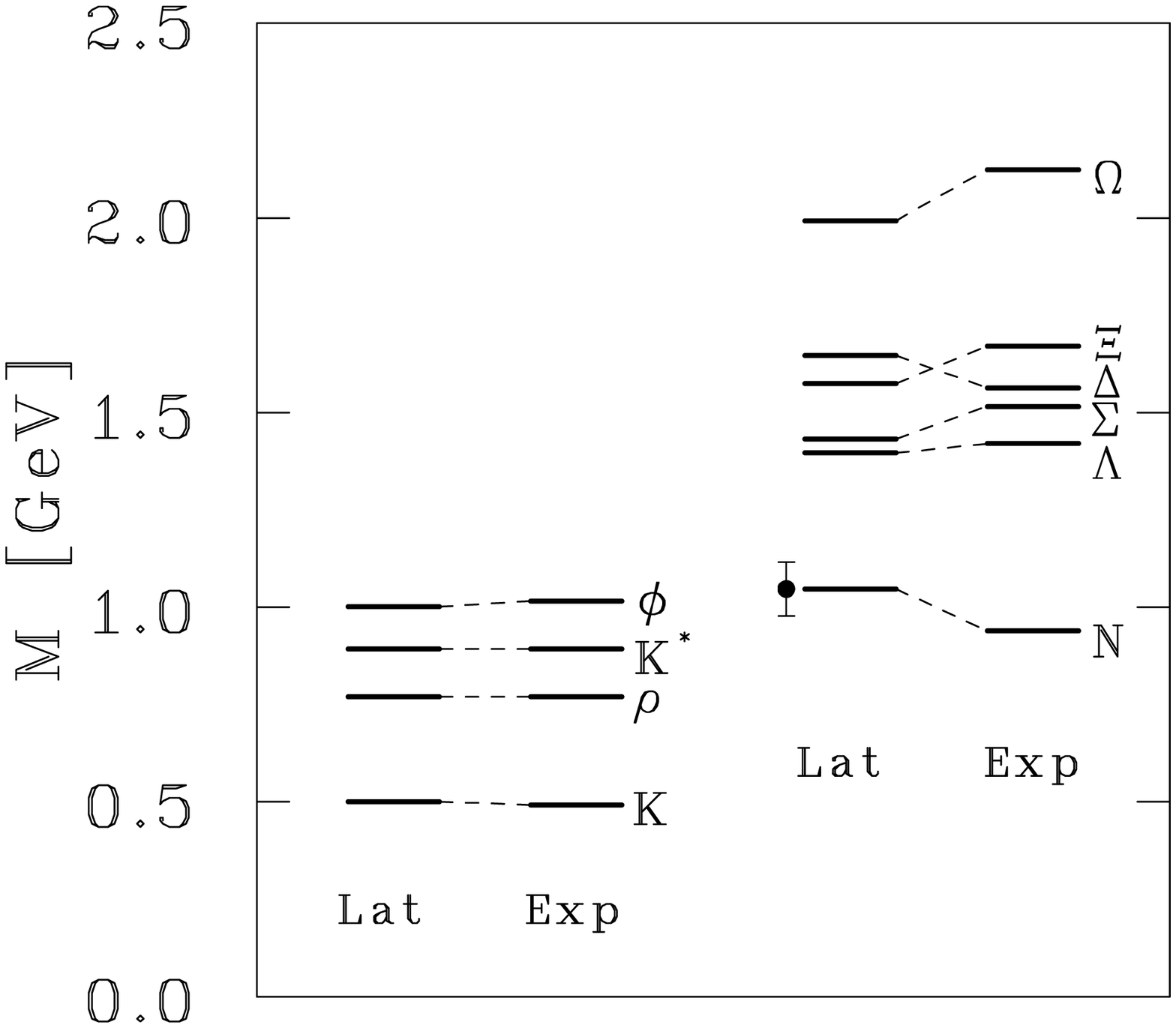,height=65mm}}
\caption{Level scheme of hadron masses from the NNN action at $\beta=6.8$
and experimental values.}
\label{fig4}\end{figure}

Calculations were carried out in quenched approximation for the NNN and
the Wilson action at matching physical lattice spacings $a_{st}$ determined
from the string tension \cite{hdt}. Two set of simulations were done at
0.4 and 0.27fm. The lattice parameters are shown in Tab.~\ref{tab1}.

Quark propagators were calculated for a range of $\kappa$ values using a
stabilized biconjugate gradient algorithm. Periodic boundary
conditions were imposed on the quark fields in spatial directions but in the
time direction a Dirichlet, or fixed, boundary condition was used. The source
position was two time steps in from the boundary in all simulations.

Meson and baryon correlators, constructed from standard local
interpolating fields, were obtained for both local and smeared
sinks with local sources. Gaussian smearing \cite{smear} was used.

Masses were calculated using an analysis procedure motivated by Bhattacharya
et al \cite{lanl}. This involves a combination of effective mass functions
from local-local and local-smeared correlators.
Except for the baryon spin-$3/2$ channel, 
it was found that compatible masses could be obtained using time averages
starting 2 or 3 time steps away from the source.

For each simulation the masses were extrapolated as a function of pion
mass to the chiral limit $M_{\pi} = 0$. The choice of extrapolation
function was either $M = M_0+cM_{\pi}^2$ or $M = M_0+cM_{\pi}^2+dM_{\pi}^3$
depending on whether the coefficient $d$ could be determined to be nonzero
within the statistical errors. The errors in masses and in mass ratios were
estimated using a bootstrap procedure with 500 samples.

In the strange quark sector we have fixed $\kappa_s$ such that $K^*/K$
equals the experimentally observed $1.8$. This was done by linear
interpolation (or extrapolation) using the two $\kappa$ values from our
set nearest to $\kappa_s$, see Tab.~\ref{tab1}.

Mass ratios extrapolated to the chiral limit are given in Tab.~\ref{tab2}.
The ratio $M_N/M_{\rho}$ sets an overall scale of baryon masses relative
to meson masses. This seems to be the quantity most effected by
discretization errors. By $0.27$fm the improved action results are fairly
close to experiment and are compatible with Wilson action calculations done
at small lattice spacing \cite{gf11}.  
 
An interesting way of looking at the relative quark mass dependence of
pseudoscalar and vector meson masses is to examine  $M_V^2-M_P^2$. 
Empirically, this quantity is approximately constant ($\approx 0.5$), a fact
which has been linked to chiral symmetry \cite{beane}.
Therefore, its behaviour in lattice QCD may
shed light on the role of chiral symmetry breaking
Wilson-like terms in the action \cite{caba}.
In Fig.~\ref{fig1} we plot the squared mass difference in the limit of zero
quark mass ($M_{\pi}=0$). At comparable lattice spacings the NNN results are
much closer to $M_V^2-M_P^2=\mbox{const}$ than the Wilson results.

In Fig.~\ref{fig2} we have compiled some results for the ratio of the lattice
spacing extracted from the string tension to the lattice spacing determined
by the ${\rho}$-meson mass \cite{lanl,gf11,jlqcd,alfor_d234,collins}.
The improved action and Wilson action results show the same qualitative
behaviour only shifted in lattice spacing by about a factor of 3.
A remarkable feature seen in Fig.~\ref{fig2}
is that different improved fermion actions exhibit a high degree of 
universality even in the non-scaling region.

If simulations done with improved actions on coarse lattices are to be useful
the results should extrapolate smoothly to the continuum limit. The ratio of
nucleon to $\rho$-meson mass has been calculated a number of times. A sample
of Wilson action results \cite{lanl,gf11,qcdpax} and results for tadpole
improved actions \cite{alfor_d234,collins} are presented in Fig.~\ref{fig3}.
It is encouraging that our improved action values at $0.27$
and $0.4$fm are compatible with the Wilson action results at smaller lattice
spacing.

In this work we have seen that, compared to the Wilson action, the tadpole
improved NNN action operates much closer to the continuum limit. Not
suprisingly, the closeness of the mass spectrum to phenomenology, see
Fig.~\ref{fig4}, confirms this conclusion.
\vspace{2ex}

It is a pleasure to thank S.R. Beane, R. Edwards, T. Klassen, F.X. Lee 
and H.D. Trottier for helpful discussions.
\vspace{1cm}


\bibliography{lat96proc}


\end{document}